
\magnification=1100
\font\bigbf=cmbx10 scaled 1400
\hfuzz 37pt
\hsize=6.1truein
\hoffset=0.2truein

\def\go{\mathrel{\raise.3ex\hbox{$>$}\mkern-14mu
             \lower0.6ex\hbox{$\sim$}}}
\def\lo{\mathrel{\raise.3ex\hbox{$<$}\mkern-14mu
             \lower0.6ex\hbox{$\sim$}}}
\def\bs{{\bf S}}
\def\bl{{\bf L}}
\def\bj{{\bf J}}
\def\bn{{\bf n}}
\def\br{{\bf r}}
\newcount\eqnumber
\def\clreqnumber{\global\eqnumber=0} \clreqnumber
\def\EQN#1#2$${\global\advance\eqnumber by1%
  \eqno\hbox{\rm({}\the\eqnumber#1)}$$\def\name{#2}\ifx\name\empty%
  \else\xdef#2{({}\the\eqnumber#1)\noexpand}\fi\ignorespaces}
\def\eqn#1$${\EQN{}{#1}$$}
\def\eqna#1$${\EQN{a}{#1}$$}
\def\eqnb#1$${\global\advance\eqnumber by-1 \EQN{b}{#1}$$}
\def\eqnc#1$${\global\advance\eqnumber by-1 \EQN{c}{#1}$$}
\def\lasteqn{{\rm({}\the\eqnumber)}}
\def\nexteqn{\advance\eqnumber by1 {\rm({}\the\eqnumber)}\advance
  \eqnumber by-1 }

\centerline{\bigbf SPIN-ORBIT INTERACTION IN}
\smallskip
\centerline{\bigbf  NEUTRON STAR/MAIN SEQUENCE BINARIES}
\smallskip
\centerline{\bigbf AND IMPLICATIONS FOR PULSAR TIMING}
\bigskip

\centerline{DONG LAI}
\centerline{Theoretical Astrophysics, 130-33, California Institute
of Technology}
\centerline{Pasadena, CA 91125}
\centerline{I: dong@tapir.caltech.edu}

\medskip
\centerline{LARS BILDSTEN}
\centerline{Department of Physics and Department of Astronomy}
\centerline{University of California, Berkeley, CA 94720}
\centerline{I: bildsten@fire.berkeley.edu}

\medskip
\centerline{VICTORIA M. KASPI$\,$\footnote{$^1$}
{Hubble Fellow.}}
\centerline{IPAC/Caltech/Jet Propulsion Laboratory, Pasadena, CA 91125}
\centerline{I: vicky@ipac.caltech.edu}

\bigskip
\centerline{\bf ABSTRACT}
\medskip

  The spin-induced quadrupole moment of a rapidly rotating star
changes the orbital dynamics in a binary system, giving rise to
advance (or regression) of periastron and precession of the orbital
plane.  We show that these effects are important in the recently
discovered radio pulsar/main sequence star binary system PSR
J0045$-$7319, and can reliably account for the observed peculiar timing
residuals. Precise measurements of the apsidal motion and orbital
plane precession can yield valuable information on the internal
structure and rotation of the star.  The detection of orbital
precession implies that the spin of the companion star is not
aligned with the orbital angular momentum, and suggests that
the supernova gave the pulsar a kick out of the original orbital
plane.  Tidal excitation of g-mode oscillations in the
PSR J0045$-$7319 system induces an orbital period change of
order $|\Delta P_{\rm orb}/P_{\rm orb}|\sim 10^{-6}$ at each periastron
passage, but the secular trend depends on the radiative damping time
of the g-modes. We also discuss the spin-orbit coupling effects for
the accreting X-ray pulsars and the other known radio pulsar/main
sequence binary, PSR B1259$-$63.

\bigskip
\noindent
{\it Subject headings:} binaries: close --- pulsars: individual:
PSR J0045$-$7319, PSR B1259$-$63 --- stars: early type --- stars:
rotation --- stars: oscillations

\bigskip
\bigskip
\centerline{To appear in {\it Astrophysical Journal\/} (October 20,
1995 issue)}

\vfil\eject
\bigskip
\centerline{\bf 1. INTRODUCTION}
\nobreak
\medskip

   The discovery of radio pulsars in binary systems with massive main
sequence stars (PSR B1259$-$63, Johnston et al.~1992, 1994 and PSR
J0045$-$7319, Kaspi et al.~1994) makes it possible to measure
hydrodynamical effects on the pulsar orbit. Indeed,
the detection of large ($\pm 30$ msec) {\it frequency-independent}
timing residuals from PSR J0045$-$7319 (Kaspi et al.~1995a) has
motivated our consideration of the dynamical effects caused by the
spin-induced quadrupole on the stellar companion. This is expected to
be the dominant effect, as massive main sequence stars are fast
rotators and typically have rotation periods on the order of days
(Jaschek \& Jaschek 1987).
We show that the spin-orbit coupling in
these systems is indeed significant, and leads to both apsidal motion
and precession of the orbital plane
that can account for the timing residuals seen in PSR J0045$-$7319.

 The information learned from pulsar/main sequence binaries will
complement the studies of double main sequence binaries, many of which
have large enough quadrupoles (either from tides or fast rotation) to
cause measurable apsidal motion in their eccentric orbits. As reviewed
by Claret \& Gimenez (1993), the apsidal motion is combined with
information on the stellar masses, radii and spins to yield the apsidal
constant, $k$, which is a dimensionless measure of the density
concentration of the stellar interior. These measurements compare
favorably with those expected from theoretical stellar structure models
(see Claret \& Gimenez 1993 for the few exceptions). The pulsar systems
considered here are most sensitive to the spin-induced quadrupole,
permitting better studies of the orientation and magnitude of the spin
of the stellar companion. Both of these quantities are important to
constrain and/or measure in the context of mass transfer prior to the
supernova and ``kicks'' during the neutron star birth.

  We start in \S 2 by explaining why the spin-induced quadrupole is
important for these systems and then summarize the full equations for
the precession and apsidal motion, correcting some previous
typographical errors. We then show in \S 3 that the induced time delay
is large and easily measurable for the two radio pulsars under
consideration, PSR J0045$-$7319 and PSR B1259$-$63 . We demonstrate
that this effect is present in the timing data of PSR J0045$-$7319.
In particular, the unusual residuals observed by Kaspi et al.~(1995a)
are the result of fitting a purely Keplerian orbit to irregularly
sampled data in which apsidal advance and orbital precession are
present.  Section 4 contains a simple estimate of the dynamical tidal
effect that might prove important. We show that tidal excitation
of high-order gravity modes in the companion of PSR J0045$-$7319
near periastron can yield significant changes in the orbital period
and eccentricity which may be measurable.
We conclude in \S 5 by outlining the
prospects for constraining the system parameters of PSR J0045$-$7319,
and the possibilities for similar measurements in accreting X-ray
pulsars.

\bigskip
\centerline{\bf 2. SPIN-ORBIT COUPLING IN BINARY SYSTEMS}
\nobreak
\medskip

The neutron star's gravitational field couples to both the
tide-induced quadrupole and the spin-induced quadrupole of the
stellar companion.  We begin by comparing the relative magnitudes of
these two effects. The tidal bulge on the stellar companion has a
height $h\sim M_pR_c^4/(M_c r^3)$, where $M_c$ and $R_c$ are the
companion mass and radius, $M_p$ is the pulsar (neutron star) mass,
and $r$ is the instantaneous center of mass separation. The tidal
quadrupole is then roughly
$Q_{\rm tide}\sim kM_cR_c^2(h/R_c)\sim kM_p R_c^2(R_c/r)^3$.
We compare this to the quadrupole induced by rotation at angular
velocity, $\Omega_s=2\pi/P_s$,
$Q_{\rm spin}\sim k M_cR_c^2(\Omega_s^2 R_c^3/GM_c)$.
At periastron, the ratio of these two is
$${Q_{\rm spin} \over Q_{\rm tide}} \sim
\left(P_{\rm orb}\over P_s\right)^2 {(M_p+M_c)\over M_p}(1-e)^3,
\eqn\qrat$$
where $P_{\rm orb}$ and $e$ are the orbital period and eccentricity.
Equation \qrat \ shows that when
$\Omega_s$ is larger than the orbital angular velocity
at periastron, $\Omega_{orb,p}=\Omega_{\rm orb} (1+e)^{1/2}/(1-e)^{3/2}$,
the spin-induced quadrupole effect is more important than the tidal effect.
Binary systems containing early-type stars (in particular those with
convective cores and radiative envelopes) with orbital periods longer
than $2-3$ days are not expected to circularize and synchronize
appreciably in a time comparable to the main sequence life span,
at least when the eccentricities are not too large
(Zahn 1977,1984).  Both PSR B1259$-$63 and PSR J0045$-$7319 are
of this type and, as we describe in \S 3, are timed with
sufficient accuracy for these effects to be measurable.

\bigskip
\centerline{\bf 2.1 Spin-Induced Distortion}
\nobreak
\medskip

  The quadrupole from the rotational distortion of the finite-sized
companion is specified by the difference between the moments of
inertia about the spin axis, $I_3$, and an orthogonal axis, $I_1$.
%
%
Claret \& Gimenez (1992) calculated the moment of inertia for slowly
rotating main sequence stars.  We incorporate their results by
introducing a parameter $\lambda$, whose value is close to unity, so
that $I_3\equiv 0.1\lambda M_c R_c^2$
(The parameter $\lambda$ also accounts for the increase of the
equatorial radius of the rotating star, which can be as large as
$3R_c/2$ near the maximum rotation rate).
The quadrupole from the rotational distortion of the star is proportional to
the spin squared.
Following the conventional definition of the apsidal motion
constant $k$ (Cowling 1938; Schwarzschild 1958), we write
$$I_3-I_1={2\over 3}k M_cR_c^2\hat\Omega_s^2, \eqn\qspinr$$
where $\hat\Omega_s$ is the dimensionless spin $\hat\Omega_s\equiv
\Omega_s/(GM_c/R_c^3)^{1/2}$ of the companion.
For stars with uniform density, $k=3/4$.
For the $M_c\approx
10M_\odot$ main-sequence star of interest here, $k\simeq 0.01$,
depending on the stellar age (Schwarzschild 1958; Claret \& Gimenez
1992).

%
%

\bigskip
\centerline{\bf 2.2 Spin-Induced Apsidal Motion and Orbital Precession}
\nobreak
\medskip

   Because of the spin-induced quadrupole moment,
the additional $1/r^3$ potential term  between the components
leads to apsidal motion (advance or regression of the longitude of
periastron) and, when the spin angular momentum
$\bs$ is not aligned with the orbital
angular momentum $\bl$, a precession of the orbital plane.  The general
expressions for the rates of apsidal motion and orbital precession have
been derived by Smarr \& Blandford (1976) and Kopal (1978) using
standard celestial mechanics perturbation theory. In the coordinates
of Figure 1, the rate of change of the dynamical longitude of
periastron (measured from the ascending node in the invariable plane
perpendicular to $\bj=\bl+\bs$), $\chi$, is given by
$$\dot\chi={3\pi
(I_3-I_1)\over M_c a^2(1-e^2)^2P_{\rm orb}} \left(1-{3\over
2}\sin^2\theta+{1\over 2}\sin 2\theta\cot\theta_c \right),
\eqn\chidot$$
where $a$ is the semi-major axis of the (relative) orbit,
$\theta$ is the angle between $\bl$ and $\bs$,
and $\theta_c$ is the angle between $\bl$ and $\bj$.
\footnote{$^2$}{This can be derived following the perturbation
procedure of \S 11-3.C of Goldstein (1980), except that here the action
variables are $J_1=2\pi L\cos\theta_c$, $J_2=2\pi L$, and note that the
perturbation Hamiltonian (11-58) depends on $\theta$ (or $i$ in
Goldstein's notation), the angle between $\bl$ and $\bs$. Note that in
Eq.~(3.10) of Smarr \& Blandford (1976), $\sin^2(\delta+\theta)$ should
be $\sin 2(\delta+\theta)$; also in Eq.~(11-60) of Goldstein (1980),
the factor $(1-e^2)^{-1}$ should be $(1-e^2)^{-2}$.}

The orbital plane precession rate
${\bf\Omega}_{\rm prec}=\dot\Phi\hat\bj$ (cf.~Fig.~1)
can be obtained more directly by
calculating the interaction torque between the stars. We have
$${\bf\Omega}_{\rm prec}
=-{3GM_p(I_3-I_1)\cos\theta\over 2a^3(1-e^2)^{3/2}}{\bj \over SL}
=-{3\pi (I_3-I_1)\over M_c a^2(1-e^2)^2P_{\rm orb}}\left({\sin\theta
\cos\theta\over\sin\theta_c}\right)\hat\bj,
\eqn\omprecc$$
where the minus sign implies that $\bl$ precesses in a direction
opposite to $\bj$, and we have used
$L=[GM_p^2M_c^2a(1-e^2)/M_t]^{1/2}$ for the Keplerian
orbital angular momentum.
In the limit of small mass ratio $M_p\ll M_c $ and
$S\equiv |\bs|\gg L\equiv |\bl|$, equations \chidot \
and \omprecc \ reduce to the rates of apsidal motion and orbital
precession for the Earth-satellite system (Goldstein 1980).
By contrast, for binary pulsars of interest in this paper,
$L\gg S$. Using the moment of inertia $I_3\equiv 0.1\lambda M_cR_c^2$
the quadrupole from equation \qspinr \ we obtain
$$\Omega_{\rm prec} \simeq -\Omega_{\rm orb}{10 kM_p\over
\lambda(M_cM_t)^{1/2}}
\left({R_c\over a}\right)^{3/2}\!\!{\hat\Omega_s\cos\theta
\over (1-e^2)^{3/2}}, \eqn\precc$$
where $M_t=M_p+M_c$.
The angle of the precession cone (i.e., the angle between
$\bl$ and $\bj$) is simply $\theta_c\simeq S\sin\theta/L$
in the $L\gg S$ limit.

The above derivation assumes the companion star
behaves as a rigid body.  One might be
concerned that fluid stars respond to external forces differently than
a rigid object (Smarr \& Blandford 1976; Papaloizou \& Pringle
1982). This is by no means a resolved issue, although it seems
unlikely that deviation from rigid body behavior is large since the
tidal distortion of the star is small compared to the rotational
distortion.  The observations of these pulsars may well provide
information on this subtlety.

\bigskip
\centerline{\bf 2.3 Effects on the Observables}
\nobreak
\medskip

 The apsidal motion and orbital plane precession change $\omega$, the
observational longitude of periastron (measured from the ascending
node in the plane of the sky), and $i$, the orbital inclination angle
(the angle between $\bl$ and $\bn$, the unit vector along the
line-of-sight).  Let $i_o$ be the angle between $\bj$ and $\bn$ (see
Fig.~1), in which case the observational angles $\omega$ and $i$ are
related to $\chi$ and $\Phi$ by: $$\eqalign{
\sin\omega &={1\over\sin i}\left[(\sin\theta_c\cos i_o
+\cos\theta_c\sin i_o\cos\Phi)\sin\chi
+\sin i_o\cos\chi\sin\Phi\right],\cr
\cos i &=-\sin\theta_c\sin i_o\cos\Phi+\cos i_o\cos\theta_c.\cr
}\eqn\angles$$
For $L\gg S $ or $\theta_c\ll 1$, these reduce to
$\omega\simeq\chi+\Phi$ and $i\simeq i_o+\theta_c\cos\Phi$.
Using equations \qspinr, \chidot \ and \omprecc, we then have
$$\dot\omega\simeq {3\pi (I_3-I_1)\over M_ca^2(1-e^2)^2P_{\rm orb}}
\left(1-{3\over 2}\sin^2\theta\right)
\simeq \Omega_{\rm orb}{kR_c^2\hat\Omega_s^2\over a^2(1-e^2)^2}
\left(1-{3\over 2}\sin^2\theta\right),
\eqn\chidotr$$
and
$${di\over dt} 
\simeq {3\pi (I_3-I_1)\over M_ca^2(1-e^2)^2P_{\rm orb}}\sin\theta
\cos\theta\sin\Phi\simeq
\Omega_{\rm orb}{kR_c^2\hat\Omega_s^2\over a^2(1-e^2)^2}
\sin\theta\cos\theta\sin\Phi.
\eqn\didt$$
Thus for typical $\theta$ and $\Phi$, the numerical value of $di/dt$
is comparable to that of $\dot\omega$. Note that the observed
apsidal motion is an advance for $\sin\theta < (2/3)^{1/2}$ and
a regression in the opposite case. In addition, twice
in one precession cycle, at $\Phi=0$ and $\pi$ (i.e., when
$\bj,\bl$ and $\bn$ lie in the same plane), the rate $di/dt$
for the change of the orbital inclination angle is identically zero.

\bigskip
\centerline{\bf 3. SPIN-ORBIT EFFECTS ON BINARY PULSAR TIMING}
\medskip

We now consider the effects of spin-orbit coupling on the timing of
the two binary pulsar systems with main-sequence-star companions.  The
orbital motion gives rise to a delay of
$T(t)=\br_p\cdot\bn/c=r_p(t)\sin\Psi(t)\sin i(t)/c$ in the pulse
arrival time, where $\br_p=\br M_c/(M_p+M_c)$ is the pulsar position
vector. The orbital phase (or true anomaly) $\Psi(t)$
measured from the observational ascending node
is given by $\Psi(t)=\Psi(t)_K+\dot\omega t$, where $\Psi(t)_K$ is the
Keplerian value. The longitude of periastron $\omega(t)$ and the
orbital inclination angle $i(t)$
vary according to equations \chidotr \ and \didt, with the
precession phase given by $\Phi(t)=\Omega_{\rm prec}t+\Phi_o$.

 We are interested in the residual $\delta T(t)\equiv
\br_p\cdot\bn/c-(\br_p\cdot\bn/c)_K$ of the time delay compared to the
Keplerian value.  For $t\ll 1/|\Omega_{\rm prec}|$ and $t\ll 1/|\dot\omega|$,
we have $\delta T=\delta T_{\rm aps}+\delta T_{\rm prec}$, where the
contributions from apsidal motion and orbital plane
precession are given by (assuming $\delta T=0$ at $t=0$)
$$\eqalign{
\delta T(t)_{\rm aps} &\simeq {r_p(t)\over c}\sin i_o\cos\Psi(t)
\,\,\dot\omega\,t,\cr
\delta T(t)_{\rm prec} &\simeq {r_p(t)\over c}\cos i_o\sin\Psi(t)
\,\,{di\over dt}\,t.\cr
}\eqn\apsid$$
Clearly, the residual resulting from the neglect of
varying $\omega$ and $i$ in the timing model increases with $t$, and
is modulated by the orbital motion
\footnote{$^3$}{Note that for very eccentric systems,
the changes in $\omega$ and $i$ mainly occur near periastron, rather
than accumulating uniformly throughout the orbital period.
We neglect this finer point in equation \apsid.}.
Equations \apsid \ are the leading
order corrections for this effect, and are valid for time scales much
less than the precession period, which is $\sim ~300$ years for PSR
J0045$-$7319 (see \S 3.1). The next order terms  are of order
$\dot\omega t$ or $(di/dt)t$ smaller and have different dependences on
$i$ and $\Psi$. The accumulation of accurate timing data over a longer
baseline will  further constrain the system.

\bigskip
\centerline{\bf 3.1 PSR J0045$-$7319}
\medskip

Radio timing observations by Kaspi et al.~(1994) of the $0.93$ s pulsar
PSR J0045$-$7319
have identified it to be in an eccentric ($e=0.8080$) 51.17 day orbit having
projected semi-major axis $a_p\sin i/c=174$ s.  They made optical observations
in the direction of the pulsar and found a
B1V star with $B=16.03$ and $V=16.19$ which
showed no evidence for emission lines.
Combining the effective temperature ($T_{\rm eff}\approx 2.4\times
10^4 \ {\rm K}$) and reddening ($A_V\simeq 0.27$) with the observed
colors gives $R_c\simeq (6.4\pm 0.5) R_\odot (d/59 \ {\rm kpc})$. The
optical radial velocity measurements of the B star fix its mass to be
$(8.8 \pm 1.8) M_\odot$ for $M_p=1.4 M_\odot$ and give $i=44\pm 5$
degrees (Bell et al.~1995). This then yields $a\simeq 126
R_\odot\simeq 20R_c$. The rotational velocity $v_{rot}\sin
\theta_{sn}\simeq 113\pm 10$ km/s (where $\theta_{sn}$ is the angle between
$\bs$ and $\bn$) obtained from spectral line broadening (Bell et
al.~1995) corresponds to $\hat\Omega_s\simeq 0.22/\sin\theta_{sn}$,
consistent with a rapid rotation near break-up,
$\hat\Omega_{s,max}\simeq 0.5$.
The ratio of $\Omega_s$ and the angular velocity at periastron
$\Omega_{orb,p}$ is $\simeq 1.1/\sin\theta_{sn}$,
confirming that the system is not synchronized.
Equations \precc, \chidotr \ and \didt \ then give
$$\eqalign{
\Omega_{\rm prec} &\simeq -1.9\times 10^{-2}\lambda^{-1}
\left({k\over 0.01}\right)\!\left({20R_c\over a}\right)^{3/2}
\!\!\!\left({\hat\Omega_s\over 0.5}\right)\cos\theta
{}~{{\rm rad}\over {\rm yr}},\cr
\dot\omega &\simeq 2.3\times 10^{-3}
\left({k\over 0.01}\right)\!\left({20R_c\over a}\right)^2
\!\!\left({\hat\Omega_s\over 0.5}\right)^2
\left(1-{3\over 2}\sin^2\theta\right)~ {{\rm rad}\over {\rm yr}},\cr
{d i\over dt} &\simeq 2.3\times 10^{-3}
\left({k\over 0.01}\right)\!\left({20R_c\over a}\right)^2
\!\!\left({\hat\Omega_s\over 0.5}\right)^2
\sin\theta\cos\theta\sin\Phi~ {{\rm rad}\over {\rm yr}}.\cr
}\eqn\jpuls$$
The timing residual $\Delta T$ accumulated in one orbit
is estimated by setting $r_p\simeq 2a_p$ and $t\simeq P_{\rm orb}/2$
in equation \apsid, giving
$$\eqalign{
\Delta T_{\rm aps} &\simeq {a_p\sin i\over c}\dot\omega P_{\rm orb}
\simeq 56\left({k\over 0.01}\right)\!\left({20R_c\over a}\right)^2
\!\!\left({\hat\Omega_s\over 0.5}\right)^2
\!\left(1-{3\over 2}\sin^2\theta\right)~{\rm ms},\cr
\Delta T_{\rm prec} &\simeq {a_p\cos i\over c}{di\over dt}P_{\rm orb}
\simeq 56\left({k\over 0.01}\right)\!\left({20R_c\over a}\right)^2
\!\!\left({\hat\Omega_s\over 0.5}\right)^2
\!\left({\cos i\over\sin i}\sin\theta\cos\theta\sin\Phi\right)
{}~{\rm ms}.\cr }\eqn\jpuls2$$
The modulated residual in just one orbit is thus comparable to the
residuals found by Kaspi et al.~(1995a).

There are other effects that can also give rise to
secular changes in $\omega$ and $i$. The standard formula for
the apsidal motion due to static tide (Cowling 1938;
Schwarzschild 1958) gives
$\dot\omega_{\rm tide}\simeq 1.3\times 10^{-4}(k/0.01)(20R_c/a)^5$
rad/yr, about an order of magnitude
smaller than the spin-induced $\dot\omega$. This secular trend is
degenerate with the spin-induced quadrupole effect.
Tidal effects do not produce changes in the orbital inclination angle.
General relativity also induces periastron advance
(Landau \& Lifshitz 1962) $\dot\omega_{\rm gr}\simeq
6.7\times 10^{-5}$ rad/yr, comparable to the tidal effect.
The geodetic precession of the companion
star's spin (Barker \& O'Connell 1975;
note that the spin angular momentum of the pulsar is much
smaller than that of the companion) results in an orbital
precession $\Omega_{\rm prec,gr}\simeq 5.9\times
10^{-6}$ rad/yr, three orders of magnitude smaller than the Newtonian
spin-induced quadrupole effect.

\bigskip
\centerline{\bf 3.2 PSR B1259$-$63}
\nobreak
\medskip

 This $47.76$ ms pulsar is in an eccentric orbit with an Be star
companion with $P_{\rm orb}=1236$ days, $e=0.8698$, and $a_p\sin
i/c=1296$ s (Johnston et al.~1992, 1994). The optical data for this
emission line star are still inadequate to place strong constraints on
the spectral type and hence radius of the star, so we will use the
fiducial values chosen by Johnston et al.~(1994), $R_c=6 R_\odot$ and
$M_c=10 M_\odot$. For $M_p=1.4M_\odot$, we obtain
$a=1091R_\odot=182R_c$. For typical values ($\lambda\simeq 1$,
$k\simeq 0.01$ and $\hat\Omega_s\simeq 0.5$) we get $\Omega_{\rm
prec}\simeq -4.1\times 10^{-5}\cos\theta$ rad/yr, $\dot\omega\simeq
2.4\times 10^{-6}(1-3/2\sin^2\theta)$ rad/yr, and $di/dt\simeq
2.4\times 10^{-6}\sin\theta\cos\theta\sin\Phi$ rad/yr. The timing
residual accumulated in one orbit is given by
$$\eqalign{
\Delta T_{\rm aps} &\simeq 10\left({k\over 0.01}\right)\!
\left({182R_c\over a}\right)^2\!\!\left({\hat\Omega_s\over 0.5}\right)^2
\!\left(1-{3\over 2}\sin^2\theta\right)~{\rm ms},\cr
\Delta T_{\rm prec} &\simeq 10\left({k\over 0.01}\right)\!
\left({182R_c\over a}\right)^2\!\!\left({\hat\Omega_s\over 0.5}\right)^2
\!\left({\cos i\over\sin i}\sin\theta\cos\theta\sin\Phi\right)
{}~{\rm ms}.\cr }\eqn\jpuls3$$
Thus $\Delta T$ can potentially affect
the timing parameters for this system as well. Indeed, using a pure
Keplerian model for the system resulted in substantial
systematic timing residuals (Johnston et al.~1994; Manchester et al.~1995).
{}From our estimate above, it is clear
that these residuals may well have a significant contribution from
the unfitted $\dot\omega$ and $di/dt$.
A full understanding of this system
is complicated by the dispersion and radio eclipse caused by
the wind from the Be star.

The tide-induced apsidal motion $\dot\omega_{\rm tide}\simeq 2.7\times
10^{-10}(k/0.01)(182R_c/a)^5$ rad/yr, as well as the general
relativistic effects, are completely negligible for this system.

Note that the expressions given in \S 2 describe the {\it secular\/}
change of the orbit, i.e., they are obtained by orbital averaging.  To
model closely-sampled timing data spanning only a few orbits, as in
PSR B1259$-$63, a numerical integration of the orbit with the
perturbing potential
may be required.

\bigskip
\centerline{\bf 3.3 Data Fitting and the Observed Residual in
PSR J0045$-$7319}
\nobreak
\medskip

    Using multi-frequency radio timing observations of PSR
J0045$-$7319, Kaspi et al.~(1995a) showed that the observed
barycentric pulse arrival times deviated significantly from those
expected from a pure Keplerian orbit. In particular, they found
significant {\it frequency-independent} timing residuals for a
$\sim$10~day duration around periastron.  These residuals showed
systematic variations on time scales of a few days after subtraction of
a Keplerian orbital fit to the entire 3.3~year data span.  The
residuals from two periastron observations separated by 11 orbits
showed very different trends, inconsistent with a simple error in any
of the Keplerian parameters. They noted that such residuals were unlike
those of any other pulsar, isolated or binary.

  We show here that the unusual PSR J0045$-$7319 residuals can be
explained by the relatively long-term dynamical effects described
above, in spite of the former's apparent short-time-scale variations.
To do this, we simulated timing observations for a binary pulsar which
has Keplerian parameters similar to those of PSR~J0045$-$7319, but in
addition has two post-Keplerian parameters, the apsidal advance
$\dot{\omega}$ and a time-variable projected semi-major axis
$\dot{x}$, where $x=a_p \sin i$. In the simulation, $x$ and $\omega$
were incremented by 0.0044~lt-s and 0.0035$^\circ$ per orbit
at periastron, which gave the best match to the observed residuals.
These numbers correspond to $\dot\omega\simeq 4.4\times 10^{-4}
{}~{\rm rad/yr}$ and $di/dt\simeq 1.8\times 10^{-4}~{\rm rad/yr}$,
consistent with equation (16). 
The actual fit to the data from PSR J0045$-$7319 and the implications
for that source in particular will be discussed elsewhere
(Kaspi et al.~1995b).

   The simulated data were then fitted with a pure Keplerian orbit, holding
all post-Keplerian parameters fixed at zero, thus simulating the
analysis done by Kaspi et al.~(1995a). The residuals for this fit are
shown in the upper panel of Figure~2.  The effect of the
post-Keplerian parameters is obvious as an orbital period modulation
of increasing amplitude relative to the fitting epoch (cf.~eq.\apsid).
The fitted Keplerian parameters are biased away from the assumed values because
of the least-squares fitting procedure; indeed the fitted solution shown
in the Figure has ``spikes'' for this reason.  Next, using
the same set of simulated arrival times, we selected those that coincide
with actual observations of PSR J0045$-$7319 at the Parkes
observatory, in order to simulate the sampling interval used by Kaspi
et al.~(1995a) in their timing analysis.  Again we fit
a simple Keplerian orbit, holding all post-Keplerian parameters
fixed at zero.  The resulting post-fit residuals are shown in the
middle panel of Figure~2.  Close-up views of the two epochs of
periastron investigated by Kaspi et al. are shown in the bottom
panels; a comparison with the data indicates that this model closely
mimics the observed trends.  Thus, we have shown that the unusual
residuals for PSR J0045$-$7319 can be explained as being a result of
fitting a purely Keplerian orbit to infrequently and irregularly
sampled data in which post-Keplerian apsidal advance and orbital plane
precession are significant.

\bigskip
\centerline{\bf 4. POSSIBLE ORBITAL PERIOD CHANGE DUE TO DYNAMICAL TIDE}
\nobreak
\medskip

The spin-induced quadrupole effects (and the standard tide-induced
apsidal motion) discussed in the previous sections
assume that at a given binary separation the star relaxes
instantaneously to the equilibrium state. However,
dynamical tidal effects associated with the excitation of
g-modes in the companion star result in energy
and angular momentum transfer between the star and
the orbit during each periastron passage, therefore
changing in the orbital period and eccentricity.
\footnote{$^4$}{Mass loss from the companion star can
in principle induce similar changes in the orbit.
However, the absence of eclipses or orbital phase dependent
DM variations constrains the ionized mass loss rate
$\dot M_c\lo 10^{-10}M_\odot~{\rm yr}^{-1}$ for PSR J0045$-$7319
(Kaspi et al.~1994,~1995a), suggesting that the dynamical importance
of the mass loss is small.
}

The energy transfer $\Delta E$ during each periastron
passage can be estimated with the formalism of Press \& Teukolsky (1977),
who considered tidal excitations for a parabolic
orbit.\footnote{$^5$}{This formalism is also reasonably accurate for
highly elliptic orbits.  This is because most of the energy transfer
takes place near periastron, where the kinetic energy and potential
energy $\sim GM_pM_c/r_{min}$ are much larger than the binding energy
$GM_pM_c/(2a)$, and the orbit resembles a parabolic trajectory.}
In the quadrupole order, this is given by
$$\Delta E\sim -{GM_p^2\over R_c}\left({R_c\over r_{min}}
\right)^6 T_2(\eta),\eqn\dedt$$
where $r_{min}$ is the binary separation at periastron,
$\eta=(M_c/M_t)^{1/2}(r_{min}/R_c)^{3/2}$ is the ratio of the time for
periastron passage and the stellar dynamical time, and the
dimensionless function $T_2(\eta)$ (defined in Press \& Teukolsky
1977) involves $l=2$ non-radial stellar oscillation modes and the
orbital trajectory. The resulting fractional change of the orbital
size is simply ${\Delta a/a}=-{\Delta E/E}$,
and the orbital period change is given by
$${|\Delta P_{\rm orb}|\over P_{\rm orb}}={3|\Delta a|\over 2a}
\sim 3{M_p\over M_c}\left({R_c\over a}\right)^5\!
(1-e)^{-6}T_2(\eta). \eqn\dpdp$$
The angular momentum transferred during periastron is
of order $\Delta L\sim \Delta E(GM_c/R_c^3)^{-1/2}$, thus
$|\Delta L/L|$ is typically negligible compared to $|\Delta E/E|$. The
resulting change in orbital eccentricity is then $\Delta
e\simeq (\Delta a/a)(1-e^2)/(2e)$.

 To estimate the importance of the dynamical tidal effects, we compare
$\Delta a/a$ with the fractional change (due to the static tide) of
$\omega$ in one orbit $\Delta\omega/(2\pi) =\dot\omega_{\rm tide}
P_{\rm orb}/(2\pi)$.
For PSR J0045$-$7319, using the parameters of \S 3.1, we have
$\eta\simeq 7.0$, and $T_2(\eta)\simeq 3.5\times 10^{-4}$ for an $n=3$
polytrope with adiabatic index $\Gamma_1=5/3$ (e.g., Lee \& Ostriker
1986), where most contributions to $T_2(\eta)$ come from g-modes of
radial order $5-9$. This gives $|\Delta a/a|\sim 0.1
\Delta\omega/(2\pi)$, implying that the perturbation due to
energy transfer is much smaller than the tide-induced apsidal motion,
and is therefore even smaller than the spin effects discussed in
\S2-3.  This estimate agrees with the more detailed study by Kumar, Ao
\& Quataert (1995).
The orbital period change due to the energy transfer is given by
$|\Delta P_{\rm orb}|/P_{\rm orb}\sim 10^{-6}$, more than an order
of magnitude larger than the measurement accuracy of $P_{\rm orb}$
attained over many orbits (Kaspi et al.~1995a).
The associated eccentricity change $|\Delta e|\sim 10^{-7}$ will
be more difficult to measure.
  For PSR B1259$-$63, again using the
parameters of \S 3.2, we have $\eta\simeq 108$, for which $T_2(\eta)$
is extremely small, and therefore the dynamical tidal effects are therefore
completely negligible\footnote{$^6$}{Kochanek (1993) has previously
discussed the dynamical tidal effects in the context of PSR
B1259$-$63. However, the observational data available then did not allow
him to draw any firm conclusion.}.

  It is important to note that the actual values and signs of $\Delta
P_{\rm orb}$ and $\Delta e$ depend on the damping of the tidally
excited g-modes. Depending on detailed stellar models, the radiative
damping time for g-modes in an early-type main sequence star
is of order $100$ years, although some g-modes
can be damped in less than a year (e.g., Unno et al.~1979).
If the damping time $t_{\rm damp}$ is much longer than the orbital
period, the energy transfer for an elliptical orbit will depend on the
phases of the oscillations of different modes, and varies (both in
magnitude and sign) from one passage to another. In this case,
equation \dpdp \ only represents the typical magnitude of $\Delta
P_{\rm orb}$ during each passage, not a steady $\dot P_{\rm orb}$;
the {\it averaged} long-term orbital decay rate is expected to be
of order $ -|\Delta P_{\rm orb}|/t_{\rm damp}$. On the other hand, if
the damping time $t_{\rm damp}$ is much shorter than the orbital period,
the energy transfer near periastron is a one-way process, i.e.,
only from the orbit to the star. Therefore
$\Delta P_{\rm orb}/P_{\rm orb}$ and $\Delta e$ represent steady
negative values, i.e., the orbit decays and circularizes.
If such a short damping timescale indeed applies to the
high order g-modes excited in PSR J0045$-$7319,
the orbital decay can be observable in the timing data.
A more accurate calculation of this effect is complicated
by the rapid rotation of the B-star, which changes the g-mode
structure when the rotation frequency is comparable to the g-mode
frequencies.


\bigskip
\centerline{\bf 5. DISCUSSION}
\nobreak
\medskip

  Smarr \& Blandford (1976) first discussed the Newtonian spin-orbit
effects in the context of PSR B1913+16. That companion turned out to be
a neutron star, which renders these Newtonian effects completely
negligible and facilitates excellent tests of general
relativity~(e.g., Taylor \& Weisberg 1989).
The recent discovery of two radio pulsar/main
sequence star binaries, PSR J0045$-$7319 and PSR B1259$-$63,
finally allows for measurements of
hydrodynamical effects in binary pulsar systems.
In this paper we have shown that the Newtonian spin-orbit
coupling is particularly important
in PSR J0045$-$7319 and PSR B1259$-$63 systems.

\bigskip
\centerline{\bf 5.1 Constraints on the Binary Systems and Evolution}
\nobreak
\medskip

When the spin-induced quadrupole is the dominant effect, measuring
$\dot\omega$ and $di/dt$ provides valuable information on the system:
(i) the sign of $\dot\omega$ constrains the spin-orbit angle $\theta$
(cf.~eq.\chidotr); (ii) the ratio of $\dot\omega$ and $di/dt$ yields a
relation between $\theta$ and $\Phi$; (iii) the angle $\theta_{sn}$ is
a function of $\theta$, $\Phi$ and $i$. Knowing $i$ and by choosing a
reasonable range of $\theta_{sn}$ (e.g., for the PSR J0045$-$7319
system: since $v_{rot}$ must be smaller than the break-up limit
$\simeq 512$ km/s, we have $|\sin\theta_{sn}| \lo 113/512=0.22$), we
obtain another relation between $\theta$ and $\Phi$; (iv) With the
angles determined from (i)-(iii), the observed values of $\dot\omega$
and $di/dt$ then constrain the stellar structure and rotation rate of
the companion. In addition, long-term timing observations will
measure various angles in addition to $\dot \omega$ and $di/dt$,
yielding additional constraints on the system parameters
(see discussion following eq.~\apsid).

   Other effects may prove important for these systems, including the
apsidal motion due to static tide and general relativity. The dynamical tide
(\S 4) may induce measurable changes in $P_{\rm orb}$ and $e$. However,
a detailed theoretical treatment of this effect is complicated by the
rapid rotation of the companion star.

   The information obtained from the systems can have important
implications for the evolution of neutron star binaries. For
example, if mass transfer in the pre-supernova binary forces $\bs$ to
be parallel to $\bl$ (a very likely scenario), then a non-zero
spin-orbit angle in the pulsar/main-sequence star system implies
that the supernova gave the neutron star a kick velocity
out of the original orbital plane. From the values of $\theta$ and $e$
of the current system (assuming that circularization and spin-orbit
alignment are not efficient in the current system), one may actually
constrain the kick velocity.
Cordes \& Wasserman (1984) discussed how
such kicks might misalign the {\it neutron star} spin from the orbital
angular momentum in PSR 1913+16, giving rise to measurable geodetic
precession.

\bigskip
\centerline{\bf 5.2 X-ray Binaries}
\nobreak
\medskip

Insight into orbital plane precession might prove important for
accreting X-ray pulsars as well. Apsidal motion has been searched for
in eccentric accreting systems. In particular, the $2\sigma$ upper
limit on the apsidal motion in the eccentric 8.9 day binary Vela X-1
(which contains a supergiant) is $\dot \omega = 1.6$ ~deg/yr (Deeter
et al.~1987). Tamura et al.~(1992) claimed a detection of apsidal
motion in the eccentric Be system 4U 0115+63 at the level $\dot
\omega=0.030 \pm 0.016 \ {\rm deg/yr}$. More recent observations of
this source with the BATSE instrument (Cominsky, Roberts \& Finger
1994) contradict the claim of Tamura et al.~(1992), find no evidence
for apsidal motion. However, these measurements could be biased by the
fact that $a_x\sin i$ was assumed constant.

  The less accurate arrival times makes the search for orbital plane
precession in the accreting X-ray pulsars more difficult than for
radio pulsars.  However, the orbital parameters are sufficiently
accurate that one can search for orbital precession through temporally
separated measurements of $a_x \sin i$, the projected size of the
orbit for the X-ray pulsar. The system 4U 0115+63
with $P_{\rm orb}=24$ days and $e=0.34$ may
experience $\Delta i\simeq 6\times 10^{-3}$~rad over a 17 year
baseline (from equation \didt \ assuming a $10 M_\odot$ main sequence
star with radius $R=6 R_\odot$, $k=0.01$ and rotating slowly enough,
$\hat \Omega_s \simeq 0.05$, or accidentally having $\sin^2\theta$
close to $2/3$ so as not to violate the upper limit on $\dot \omega $
from Cominsky et al. 1994).  This is a factor of 6 larger than the
fractional error on $a_x\sin i$ measured in 1978 (Rappaport et
al. 1978) and may well allow for a very sensitive measurement of the
rotation of the Be star in this system.

\bigskip
\bigskip

We thank Roger Blandford and Sterl Phinney for helpful discussion.
This work has been supported in part by NASA Grants
NAGW-2394 to Caltech and NAG5-2819 to U.C. Berkeley. L. B. was
supported as a Compton Fellow at Caltech through NASA grant
NAG5-2666 during part of this work.
V.M.K. thanks Dave Van Buren and is supported by a Hubble
Fellowship through grant number HF-1061.01-04A from the Space
Telescope Science Institute, which is operated by the Association
of Universities for Research in Astronomy, Inc., under NASA
contract NAS5-26555.  Research by V.M.K. was carried out at the
Jet Propulsion Laboratory, California Institute of Technology,
under a contract with the National Aeronautics and Space
Administration.

\vfill\eject
\centerline{\bf REFERENCES}
\nobreak
\medskip
\def\bysame{\hbox to 50pt{\leaders\hrule height 2.4pt depth -2pt\hfill .\ }}
\def\hi{\noindent \hangindent=2.5em}

\hi{
Barker, B.~M., \& O'Connell, R.~F. 1975, ApJ, 199, L25}

\hi{
Bell, J. F., Bessell, M. S., Stappers, B. W., Bailes, M. \& Kaspi, V.
1995, ApJL, in press}

\hi{
Claret, A., \& Gimenez, A. 1992, A\&A Suppl., 96, 255}

\hi{
\bysame 1993, A\&A, 277, 487}

\hi{
Cominsky, L., Roberts, M. \& Finger, M. H. 1994 in {\it The Second
Compton Symposium}, ed. C. E. Fichtel, N. Gehrels, \& J. P. Norris
(AIP: New York) p. 294

\hi{
Cordes, J. M. \& Wasserman, I. 1984, ApJ, 279, 798}

\hi{
Cowling, T.~G. 1938, MNRAS, 98, 734}

\hi{
Deeter, J.~E., Boynton, P.~E., Lamb, F.~K., \& Zylstra, G. 1987, ApJ,
314, 634}

\hi{
Goldstein, H. 1980, Classical Mechanics, 2nd Ed. (Addison-Wesley),
p.~512}

\hi{
Jaschek, C., \& Jaschek, M. 1987, The Classification of
Stars (Cambridge Univ. Press: New York)}

\hi{
Johnston, S. et al. 1992, ApJ, 387, L37}

\hi{
Johnston, S. et al. 1994, MNRAS, 268, 430}

\hi{
Kaspi, V.~M., et al. 1994, ApJ, 423, L43}

\hi{
Kaspi, V.~M., Manchester, R.~N., Bailes, M. \& Bell, J.~F. 1995a, Compact
Stars in Binaries, IAU Symp. 165, ed. J. van Paradijs, E.~P.~J.~ van~den
Heuvel, \& E. Kuulkers (Kluwer Dordrecht), in press.}

\hi{
Kaspi, V.~M., et al. 1995b, in preparation.}

\hi{
Kochanek, C.~S. 1993, ApJ, 406, 638}

\hi{
Kopal, Z. 1978, Dynamics of Close Binary Systems (D. Reidel Pub.:
Dordreicht, Holland), p 232}

\hi{
Kumar, P., Ao, C.~O., \& Quataert, E.~J. 1995, ApJ, in press}

\hi{
Landau, L.~D., \& Lifshitz, E.~M. 1962, Classical Theory of
Fields (Oxford: Pergamon)}

\hi{
Lee, H.~M., \& Ostriker, J.~P. 1986, ApJ, 310, 176}

\hi{
Manchester, R.~N., et al. 1995, ApJL, in press}

\hi{
Papaloizou, J.~C.~B., \& Pringle, J.~E. 1982, MNRAS, 200, 49}

\hi{
Press, W.~H., \& Teukolsky, S.~A. 1977, ApJ, 213, 183}

\hi{
Rappaport, S., Clark, G. W., Cominsky, L., Joss, P. C. \& Li, F. 1978,
ApJ, 224, L1}

\hi{
Schwarzschild, M. 1958, Structure and Evolution of the Stars
(Princeton University Press)}


\hi{
Smarr, L.~L., \& Blandford, R.~D. 1976, ApJ, 207, 574}

\hi{
Tamura, K., Tsunemi, H., Kitamoto, S., Hayashida, K. \&
Nagase, F. 1992, ApJ, 389, 676}

\hi{
Taylor, J.~H. \& Weisberg, J. ~M. 1989, ApJ, 345, 434}

\hi{
Unno, W., Osaki, Y., Ando, H., \& Shibahashi, H. 1979,
Nonradial Oscillations of Stars (Univ. of Tokyo Press), pp.~205-217.}

\hi{
Zahn, J.-P. 1977, A\&A, 57, 383}

\hi{
\bysame 1984, in Observational Tests of the Stellar Evolution
Theory, eds. A. Maeder and A. Renzini (Dordrecht: Reidel), p.~379.}

\vfil\eject
\centerline{\bf Figure Captions}
\nobreak
\medskip

\noindent
{\bf FIG.~1}.---
Binary geometry and the definitions of different angles.
The invariable plane (XY) is perpendicular to the total
angular momentum vector $\bj=\bl+\bs$, and the line-of-sight
unit vector $\bn$ lies in the YZ plane, making an angle $i_o$
with $\bj$. The inclination of the orbit with respect to the
invariable plane is $\theta_c$ (which is also the precession
cone angle of $\bl$ around $\bj$), while the orbital inclination
with respect to the plane of the sky is $i$. The angle between
$\bl$ and $\bs$ is $\theta$. The orbital plane intersects the
invariable plane at ascending node A, with a longitude
$\Phi$ measured in the invariable plane ($\Phi$ is also
the phase of the orbital plane precession).
The dynamical longitude of periastron (point P), measured
in the orbital plane, is $\chi$. The observational
longitude of periastron is $\omega$ (not shown in the Figure).

\bigskip
\noindent
{\bf FIG.~2}.--- Simulated pulse arrival times for PSR J0045$-$7319.
The upper panel shows the residuals from a fit of a pure Keplerian
orbit to fake pulse arrival time data (``measured'' daily for four
years) for a binary pulsar like PSR J0045$-$7319 in which apsidal
advance and orbital plane precession are important.  The middle panel
shows the residuals from a fit of this fake data sampled only at those
epochs coinciding with actual observations of the pulsar.  The two
vertical solid lines show epochs of periastron separated by 11 orbits.
The lower panels show close-ups of those epochs; the observed trends
in the residuals are similar to those observed by Kaspi et
al.~(1995a).

\end